\documentclass[%
 reprint,
superscriptaddress,
 amsmath,amssymb,
 aps,
]{revtex4-2}
\usepackage{physics}
\usepackage[dvipsnames]{xcolor}
\usepackage{graphicx}
\usepackage{dcolumn}
\usepackage{bm}
\usepackage{hyperref}

\begin{document}

\preprint{APS/123-QED}

\title{Blueprint for efficient nuclear spin characterization with color center}

\author{Majid Zahedian}
 \affiliation{3rd Physical Institute University of Stuttgart, Germany}
 \affiliation{Institute of Quantum Electronics, ETH Z\"urich, Switzerland}

\author{Vadim Vorobyov}
\email[]{v.vorobyov@pi3.uni-stuttgart.de}
 \affiliation{3rd Physical Institute University of Stuttgart, Germany}  

\author{J\"org Wrachtrup}
 \affiliation{3rd Physical Institute University of Stuttgart, Germany}
 \affiliation{Max Planck Institute for solid state research, Stuttgart, Germany}

\date{\today}

\begin{abstract}
Nuclear spins in solids offer a promising avenue for developing scalable quantum hardware. 
Leveraging nearby single-color centers, these spins can be efficiently addressed at the single-site level through spin resonance.
However, characterising individual nuclear spins is quite cumbersome since the characterisation protocols may differ depending on the strength of the hyperfine coupling, necessitating tailored approaches and experimental conditions.
While modified electron spin Hahn echoes like CPMG and XY8 pulse sequences are commonly employed, they encounter significant limitations in scenarios involving spin-1/2 systems, strongly coupled spins, or nuclear spin baths comprising distinct isotopes.
Here, we present a more straightforward approach for determining the hyperfine interactions among each nuclear and the electron spin. 
This method holds promise across diverse platforms, especially for emerging S=1/2 group IV defects in diamond (e.g., SiV, GeV, SnV, PbV) or silicon (T-centre, P-donors). 
We provide a theoretical framework and adapt it for color-centers exhibiting various spins. Through simulations conducted on nuclear spin clusters, we evaluate different protocols and compare their performance using the Fisher information matrix and Cramer Rao bounds.
\end{abstract}

\maketitle
\section{Introduction and background}
Optically active defects in solids, known as color centers, have been utilized in various quantum applications \cite{awschalom2018quantum}, including quantum networks \cite{pompili2021realization}, quantum sensing \cite{degen2017quantum, meinel2023high}, and quantum registers \cite{bradley2019ten, vorobyov2021quantum, hesselmeier2023measuring, hesselmeier2024high, reiner2024high}. 

\begin{figure*}
	\begin{center}
		\includegraphics[width=0.9\textwidth]{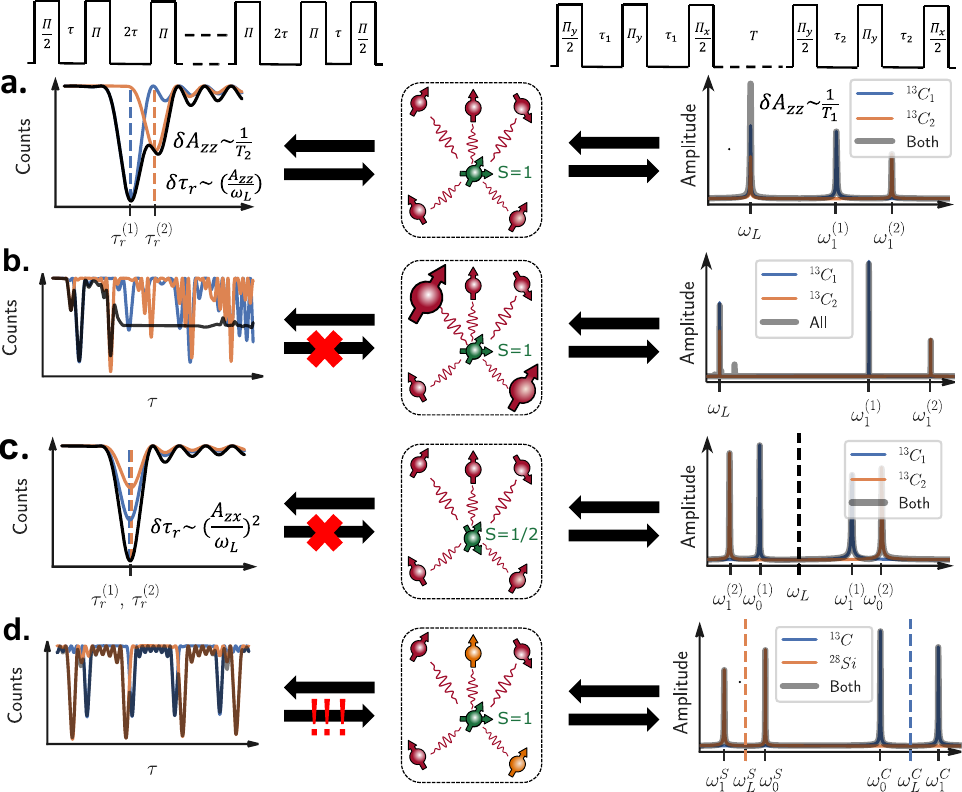}
		\caption{Comparing DD and five pulse ESEEM as two nuclear spin characterization methods in different systems. a. Electron spin one system: In the DD signal, each nuclear spin exhibits a distinct resonance time up to the first order of $A_{zz}$; the accuracy of the obtained hyperfine interaction is limited to $T_2$ of the electron spin. In the Fourier transform of the ESEEM signal, the two resonance frequencies of each spin are present; the accuracy of the obtained hyperfine interaction is limited to $T_1$ of the electron spin. b. Electron spin 1 system with two strongly coupled nuclear spins and many weakly coupled spins: No resonance time is detectable since the required condition for DD ($\omega_L \gg A_{zz}, A_{zx}$) does not hold. However, resonant frequencies are still obtainable in the ESEEM signal. c. Electron spin 1/2 system: Nuclear spins have more or less the same resonance time as it depends on the second order of hyperfine coupling. However, the resonant frequency of each spin is distinguishable in the ESEEM signal. d. System containing two different nuclear spin species: Two nuclear spin baths can interfere in the DD signal, making the signal analysis more challenging. However, the species can be inferred from the ESEEM signal since species have different Larmor frequencies.}
		\label{fig:1}
	\end{center}
\end{figure*}

Each center contains an electron spin that can be directly controlled using microwaves, and it can be initialized and read out through optical excitation. 
The potential hyperfine interactions with a bath of numerous nuclear spins are crucial for diverse applications, as they offer long-lived quantum memories and enable the creation of an optically accessible nuclear spin qubit register.
To implement efficient quantum control for these memories, it is imperative to have precise knowledge of the full Hamiltonian governing the register \cite{vorobyov2023transition,bradley2022robust}.
However, characterizing the hyperfine coupling of nuclear spins can be both challenging and time-consuming \cite{abobeih2019atomic, van2023mapping}. 
Traditionally, nuclear spin characterization is accomplished through Optically Detected Magnetic Resonance (ODMR) \cite{Dreau:2012aa,hesselmeier2023measuring}, but the spectroscopic resolution is constrained by the $1/T_2^\ast$ of the electron spin. 
To overcome this limitation, Hahn-Echo-type sequences have been employed to refocus the electron spin and extend its coherence time \cite{childress2006coherent}, thereby improving the resolution to $1/T_2$. 
This method could be applied to defects with a specific configuration of nuclear spins, particularly those with a certain relation to the magnetic field and the coupling of the nuclear spins, such as weakly coupled nuclear spins.
Also, the protocols can only be applied to electron spins with a specific spin multiplicity.
Depending on the applied protocol for the nuclear spins characterisation, the number of identifiable nuclear spin can be different.
Electron Spin Echo Envelop Modulation (ESEEM) types of sequences allow access to the highest number of nuclear spins, as the spectroscopy resolution is limited to the longitudinal relaxation time ($T_1$) of the system. (see fig. \ref{fig:1}) \cite{laraoui2013high, vorobyov2022addressing}.
In particular, we consider the following common examples of nuclear spin registers that are ubiquitous in applications. First, the NV-like case with $S=1$ and weakly coupled nuclear spins, operating at high magnetic fields where $\gamma_n B \gg A_{zx}$ which has been extensively studied \cite{abobeih2019atomic, taminiau2012detection, Zhao:2014aa}.
The capability to observe narrow peaks with analytically predictable positions and contrasts enables the solution of the inverse problem concerning the characterization of the coupling between nuclear and electron spins. 
The second case (as depicted in Fig. \ref{fig:1}b) corresponds to a scenario similar to the first, but with the presence of a strongly coupled nuclear spin, which obstructs the observation of weakly coupled nuclear spins.
This case is particularly relevant in situations where, in addition to the weakly coupled register, a strongly coupled nuclear spin is utilized, for instance, for repetitive enhancement of readout \cite{neumann2010single}.
This enhancement is achieved by exploiting the strongly coupled ancilla, which is discernible in electron spin resonance spectra.
The third case (Fig. \ref{fig:1}c) pertains to the $S=1/2$ scenario, where the lack of offset in the average evolution of the nuclear spin results in its dynamics weakly depending on $A_{zz}$ and only the second order dependence on $A_{zx}$.
Consequently, this leads to a limited ability to distinguish between multiple nuclear spins and a lack of individual addressability of nuclear spins.
Lastly, in the case of other material host platforms, a bispecies nuclear spin bath may be observed, where resonance peaks corresponding to different Larmor frequencies of different species further complicate the reconstruction of the Hamiltonian.
For unambiguous reconstruction of the interactions, additional measurements, such as those at different magnetic field strengths, would be required.
While double resonance methods \cite{Bradley:2019aa,zaiser2016enhancing} hold large promise for nuclear spin spectroscopy, they require additional experimental hardware, and often are not available. 
Thus, in this work, we investigate the limits of electron spin driven schemes, in particular a correlation-type sequences, which could serve as a general framework and hold promise for the reconstruction of the interactions in all of the aforementioned cases.

\section{Results}
\label{sec:theory}
\begin{figure*}
	\begin{center}
		\includegraphics[width=0.9\textwidth]{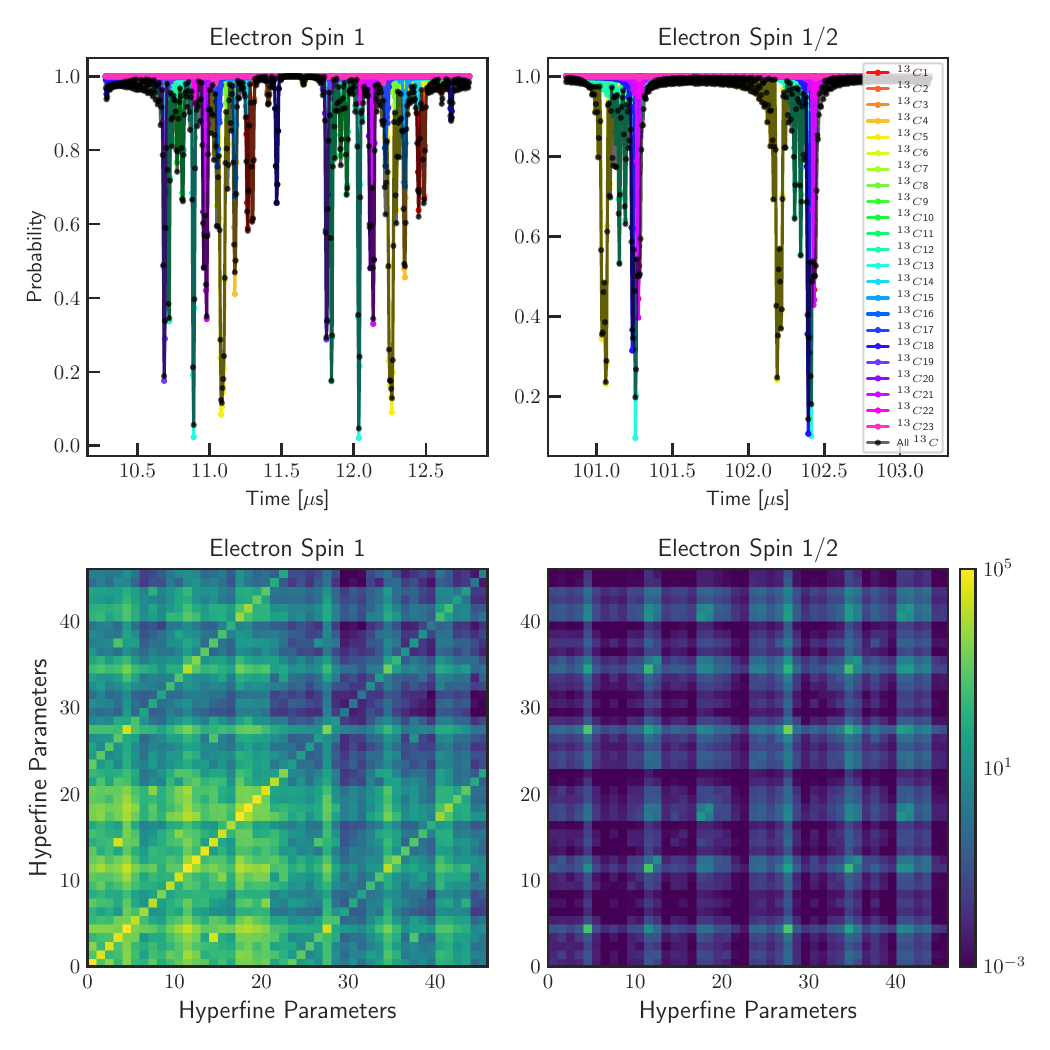}
		\caption{Simulated DD signal with 64 $\pi$-pulses for the register reported in  \cite{abobeih2019atomic} consisting of 23 nuclear spins, assuming the electron spin is a. spin $S=1$ system b. spin $S=1/2$ system. Simulated Fisher information matrix for the case of the electron spin c. spin $S=1$ system d. spin $S=1/2$ system. The first 23 hyperfine parameters are $A_{zz}$ and the second 23 parameters are $A_{zx}$ of the nuclear spin register.}
		\label{fig:3}
	\end{center}
\end{figure*}
We consider a central spin system of \textit{non-interacting} nuclear spins $I=1/2$ coupled to a central electron spin $\boldsymbol{S}$.
The central electron spin is manipulated resonantly with microwave pulses, which transfer the population between the two sublevels denoted as $m_s=s_0$ and $m_s = s_1$.
These two spin sublevels are separated due to the Zeeman effect and/or zero-field splitting, forming a two-level subsystem with an energy splitting of $\omega_a$.
The Hamiltonian of the system in the secular approximation and in the rotating frame of the applied microwave $\omega_{mw}$ can be written as:
\begin{align}
	H = \Delta S_z + \sum_k (\omega_L^{(k)}  + A_{zz}^{(k)} S_z) I_z^{(k)} + A_{zx}^{(k)} S_z I_x^{(k)}
	\label{eq:H}
\end{align}
Where $\Delta = \omega_a - \omega_{mw}$ is the detuning, $\omega_L^{(k)} =\gamma_n^{(k)} B$ is the nuclear Rabi frequency with $\gamma_n^{(k)}$ being the nuclear spin gyromagnetic ratio and $B$ the external magnetic field, and $A_{zz}$ and $A_{zx}$ are the parallel and perpendicular secular components of the hyperfine tensor, respectively.
The Hamiltonian is diagonal in the electron spin subspace, thus the Hamiltonian for the nuclear spins could be rewritten in the electron spin subdomains $s_i$ ($i=0,1$).
The nuclear spin Hamiltonian could be solved for eigenenergies, which determine the precession frequencies of $\omega_i = \sqrt{(\omega_L + s_i A_{zz})^2+ (s_i A_{zx})^2}$, $i=0,1$ along the eigenvector axes $\vec{n}_i =(\frac{s_i A{zx}}{\omega_i}, 0, \frac{\omega_L + s_i A_{zz}} {\omega_i})$.
It is assumed that the pulse duration $t_p$ is short enough ($\omega_L t_p \ll 1$) that the nuclear spin dynamics during this time is negligible.
For completeness, we start our analysis with the simplest characterization sequence, which is the electron spin Ramsey sequence (free induction decay of the electron).
The Ramsey signal can be obtained as follows:
\begin{gather}
	\expval{\sigma_z}_{Ram} = \cos(\Delta \tau) \prod_{j=1}^{n} \Big[ \cos(\frac{\omega_0 \tau}{2}) \cos(\frac{\omega_1 \tau}{2})+ \notag \\
	(\vec{n_0}\cdot \vec{n_1})\sin(\frac{\omega_0 \tau}{2}) \sin(\frac{\omega_1 \tau}{2}) \Big]^{(j)}
	\textcolor{gray}{e^{-(\frac{\tau}{T_2^\ast})^m}}
\end{gather}	
The exponential decay is introduced to note the Ramsey signal decays with the electron spin $T_2^\ast$ since the formalism does not account for any relaxation processes. The Ramsey sequence is sensitive to detuning since the electron spin is not refocused; however, it is possible to sense nuclear spins with vanishing perpendicular hyperfine coupling.\\

To extend the relaxation time and access more nuclei, one can insert a $\pi$ pulse between the Ramsey sequence, creating a Hahn echo sequence, and obtain the following signal:
\begin{gather}
	\expval{\sigma_z}_{HE} = \prod_{j=1}^n \Big[1 - 2 k^2 
	\sin^2(\frac{\omega_0 \tau}{2}) \sin^2(\frac{\omega_1 \tau}{2})\Big]^{(j)} \textcolor{gray}{ e^{-\frac{2\tau}{T_2}}}
\end{gather}
Where $k = \frac{(s_1-s_0)\omega_L A_{zx}}{\omega_0 \omega_1}$ is the modulation amplitude of each nuclear spin.
Without any nuclear spin, the Hahn Echo sequence creates an electron spin echo signal with an envelope of stretched exponential decay with $T_2^{HE}$.
However, in the presence of nuclear spins, the electron spin echo envelope will be modulated due to interaction with the nuclei.
Hence, this sequence is also called Electron Spin Echo Envelope Modulation or simply ESEEM.
Even though the coherence time is increased, distinguishing the effect of different nuclear spins in the total signal is very complicated.
The Hahn Echo sequence can be used for defects that contain a few strongly coupled nuclei.
In order to differentiate the resonance frequency of each nuclear spin or to sharpen the oscillations, one can add more $\pi$ pulses, creating the so-called Dynamical Decoupling (DD) sequence to separate the resonance condition for individual nuclear spin. For the even number of $\pi$ pulses, $N$, The DD signal can be obtained (neglecting decoherence terms):
\begin{gather} \label{eq:DD}
	\expval{\sigma_z}_{DD} =
	\prod_{j=1}^n \Big[1- 2 k^2  
	\sin^2(\frac{\omega_0 \tau}{2}) \sin^2(\frac{\omega_1 \tau}{2}) \frac{\sin^2(\frac{N}{2} \theta)}{\cos^2(\frac{1}{2}\theta)}  \Big]^{(j)} \textcolor{gray}{ e^{-\frac{2N\tau}{T_2(N)}}} 
\end{gather}
\normalsize
Where
\begin{gather}
	\theta=  \arccos\Big[\cos(\omega_0 \tau) \cos(\omega_1 \tau)- \vec{n}_0 \cdot \vec{n}_1 sin(\omega_0 \tau) \sin(\omega_1 \tau) \Big]
\end{gather}
Where $\vec{n}_0 \cdot \vec{n}_1 = \frac{\omega_L^2+(s_0+s_1)A_{zz}+s_0 s_1 (A_{zz}^2+A_{zx}^2)}{\omega_0 \omega_1}$.
At this point, it should be clear that the number of pulses can be used as an additional parameter to modify the modulation depth of each nuclear spin.
A detailed investigation of the obtained DD signal reveals that by sweeping the time interval between pulses, an exponential decay with the rate of $1/T_2$ is observed, except for some resonance times where the electron spin becomes entangled with a particular nuclear spin, resulting in a sharp drop in the signal.
This resonance time can be obtained assuming $\omega_L \gg A_{zz}, A_{zx}$ is satisfied: 
\begin{gather} \label{eq:resonance}
	\tau_p \approx  \frac{(2p+1) \pi}{\omega_0 + \omega_1} \approx \frac{(2p+1)\pi}{2\omega_L(1+\frac{s_0+s_1}{2} \frac{A_{zz}}{\omega_L}+ \frac{s_0^2+s_1^2}{4}\frac{A_{zx}^2}{\omega_L^2})}
\end{gather}
Experimental observation of the resonance times provides valuable information about the nuclear spins.
For instance, in the case of an NV center, the second-order term $\frac{A_{zx}^2}{\omega_L^2}$ can be neglected, allowing for the direct determination of the $A_{zz}$ hyperfine component from the resonances.
To proceed with the characterization, Eq. \ref{eq:DD} needs to be further simplified, requiring stronger assumptions.
Focusing on weakly coupled nuclear spins or high magnetic fields, the multiplication over all nuclear spins in Eq. \ref{eq:DD} can be approximated by a summation rule, neglecting the higher-order cross-resonance terms:
\small
\begin{gather}
	\expval{\sigma_z}_{DD} \approx 1 -2
	\sum_{j=1}^n \Big[ k^2 	 \sin^2(\frac{\omega_0 \tau}{2}) \sin^2(\frac{\omega_1 \tau}{2}) \frac{\sin^2(\frac{N}{2} \theta)}{\cos^2(\frac{1}{2}\theta)}  \Big]^{(j)} \textcolor{gray}{ e^{-\frac{2N\tau}{T_2(N)}}}
	\label{eq:approx_DD}
\end{gather}
\normalsize
Assuming non-overlapping resonance times (which does not hold true for spin 1/2 systems) and expanding the signal around the j-th nuclear spin resonance ($\tau = \tau_p^{(j)}+\delta \tau$) up to the first order of $\delta \tau$ results in approximating each drop with a Lorentzian function:
\small
\begin{gather}
	\expval{\sigma_z}_{DD} (\delta \tau ) \approx 1 -2\sin^2(\frac{N}{2}\frac{(s_1-s_0)A_{zx}}{\omega_L}) \frac{(\frac{(s_1-s_0)A_{zx}}{2\omega_L^2})^2}{{\delta \tau}^2 +(\frac{(s_1-s_0)A_{zx}}{2\omega_L^2})^2}
\end{gather}
\normalsize
This simplification allows for the determination of the parallel component of the hyperfine interaction for each nuclear spin, possibly in two ways. First, a Lorentzian with a width of $\frac{(s_1-s_0) A_{zx}}{\omega_L^2}$ can be fitted to a dip to identify $A_{zx}$. Second, by keeping track of the minima of a dip while varying the number of pulses, the periodic function $1 - 2 \sin^2\big(\frac{N}{2}\frac{(s_1-s_0) A_{zx}}{\omega_L}\big)$ can be fitted to obtain $A_{zx}$.

Although the Dynamical Decoupling (DD) sequence is well understood and used to characterize nuclear spins, it is not applicable to all systems.
First, DD only works for defects in high magnetic fields and low hyperfine coupling ($\omega_L \gg A_{zz}, A_{zx}$).
Second, the signal is limited to the electron spin $T_2$, which only gradually approaches $T_1$.
Third, signal analysis is rather complicated and time-consuming; one has to collect enough data to ensure that no two nuclear spins overlap in the signal.
Fourth, this sequence does not work for spin 1/2 systems, such as group IV defects in diamond, since the parallel component of the hyperfine coupling cancels their effect on the first order electron spin signal.
Hence, the resonance signal depends on the second order of the perpendicular hyperfine component.
\begin{figure*}[htbp]
	\begin{center}
		\includegraphics[width=0.9\textwidth]{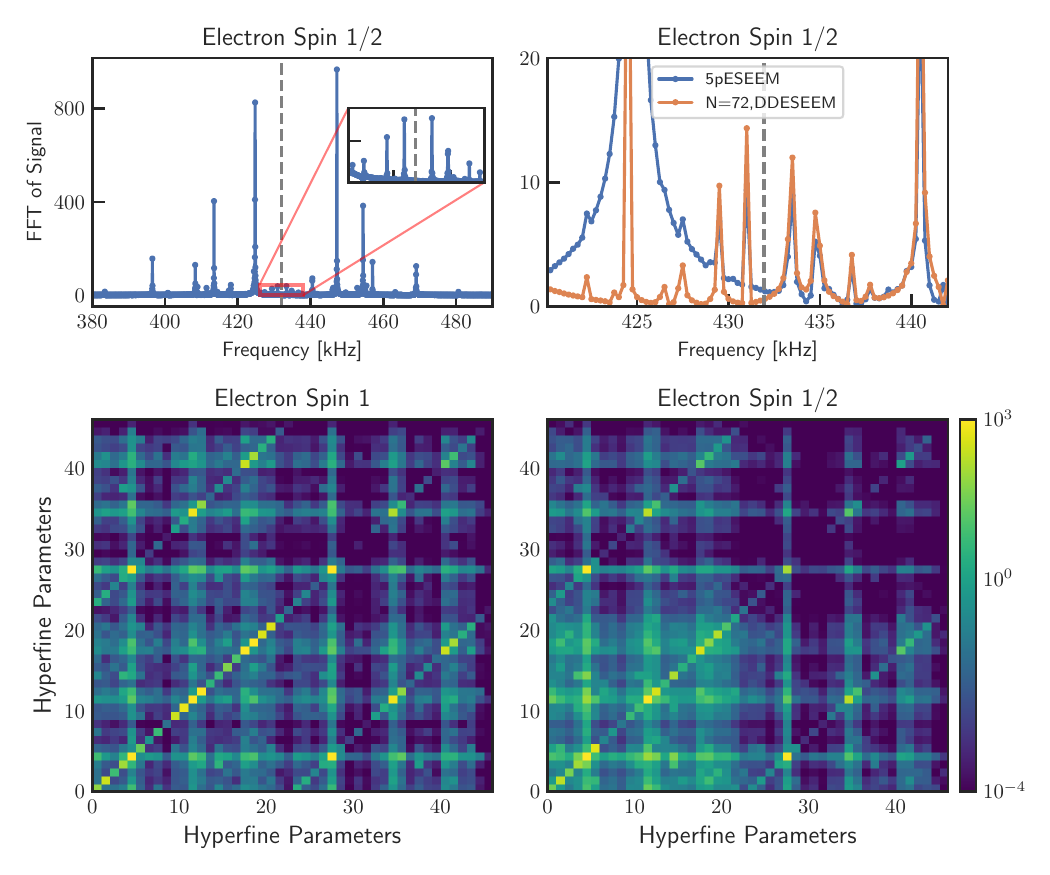}
		\caption{Simulated FFT of ESEEM signal with a zoom into weakly coupled nuclear spins for the register reported \cite{abobeih2019atomic} consist of 23 nuclear spins, assuming the electron spin $S=1/2$ system. b. Enhancing the sensitivity to weakly coupled spins by increasing number of $\pi$-pulses to $N=72$ pulses in each entangling period. $\tau$ is set to the Larmor bright spot. Simulated Fisher information for the case of the electron spin c. $S=1$ d. $S=1/2$ system. The first 23 hyperfine parameters are $A_{zz}$ and the second 23 parameters are $A_{zx}$ of the nuclear spin register.}
		\label{fig:ddeseem}
	\end{center}
\end{figure*}

In this work, we consider the 23-nuclear spin cluster characterized previously \cite{abobeih2019atomic} as a realistic case for our comparison.
First, we apply this to the dynamical decoupling sequence and visualize the obtained results. Fig \ref{fig:3} compares the DD signal for the same nuclear spin register but different defects in diamond.
An intuitive description of the sensitivity of a sequence to the variation of hyperfine coupling could be estimated by considering the width and sensitivity of the position of the resonances.
First, we consider the case of an NV center in diamond. The width of a dip is $\delta \tau = \frac{A_{zx}}{\omega_L^2}$. Taking the derivative of the resonance time, we define the sensitivity of the longitudinal hyperfine coupling, using Eq. \ref{eq:resonance} to simplify:
\begin{gather}
	\delta A_{zz} \equiv \frac{\delta \tau}{\abs{\frac{\partial \tau_p}{\partial A_{zz}}}} = \frac{2}{\tau_p} \frac{A_{zx}}{\omega_L} \ge \frac{4}{T_2^{HE}} \frac{A_{zx}}{\omega_L}
\end{gather}
If we assume typical values of $T_2^{HE} = 100$ $\mu$s and $\omega_L=500$ kHz, a weakly coupled nuclear spin $A_{zx}=5$ kHz can be distinguished from another weakly coupled nuclear spin with $\delta A_{zz} = 400$ Hz.
For $S=1/2$, such as group IV defects in diamond, the resonance times are sensitive to the second order of $A_{zx}$ and show no sensitivity to $A_{zz}$.
Hence, even at large $\tau$, the resonances related to nuclear spins will not be well separated.
We define the sensitivity for the spin one-half system as follows:
\begin{gather}
	\delta A_{zx} \equiv \frac{\delta \tau}{\abs{\frac{\partial \tau_k}{\partial A_{zx}}}} = \frac{4}{\tau_k} \ge \frac{8}{T_2^{HE}}
\end{gather}
Assuming typical values of $T_2^{HE} = 100$ $\mu$s, a nuclear spin can be distinguished from another one if their transverse hyperfine coupling is separated by $\delta A_{zx} = 80$ kHz, which is quite inaccurate and approximately 160 times worse than for $S=1$.
\begin{figure*}
	\begin{center}
		\includegraphics[width=0.9\textwidth]{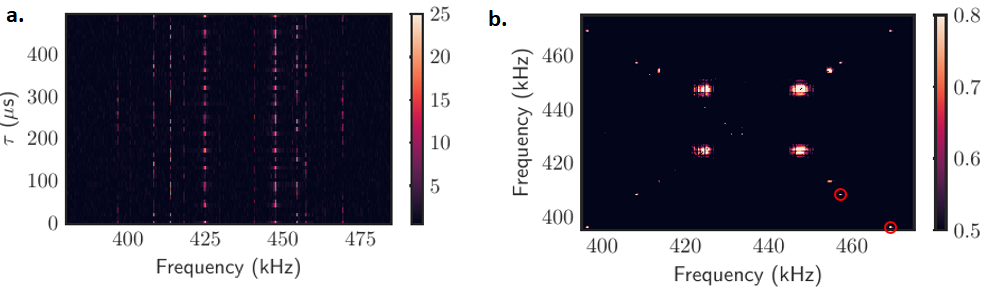}
		\caption{a. FFT of ESEEM signal for $\tau$ from 10 $\mu$s to 500 $\mu$s with a step of 10 $\mu$s b. 2D correlation of each frequency for different $\tau$.}
		\label{fig:4}
	\end{center}
\end{figure*}

To quantify the difference in sensitivity to estimating the hyperfine parameters, we perform calculations of the Fisher Information Matrix (FIM) for the 23-nuclear spin cluster, with 23 $A_{zz}$ and 23 $A_{zx}$ parameters (see Fig. \ref{fig:3}c,d).
The Fisher Information Matrix is estimated for the probability to measure state $|0\rangle$ and reads as: 
\begin{equation}
	F_{ij}(\textbf{A})=\sum_\tau \frac{\partial p(0, \tau, \textbf{A})}{\partial A_i}\frac{\partial p(0, \tau, \textbf{A})}{ \partial A_j} \frac{1}{p (1-p)}.
\end{equation}
For the typical case of an NV center with $S=1$, since all the nuclear spin resonances are clearly resolved, the Fisher Information Matrix takes a diagonal shape, revealing low covariances between the various nuclear spin resonances.
There is crosstalk between $A_{zz}$ and $A_{zx}$ for a nuclear spin, as the position of the peaks depends on both values.
The Fisher Information provides a bound for the precision of parameter estimation, known as the Cramér–Rao bound: 
\begin{equation}
	\delta A^2 \ge \frac{1}{N} F(A)^{-1}
\end{equation}
The striking difference to the $S=1$ Fisher Information Matrix is the behavior of the defect with $S=1/2$.
First of all, for both spin $S=1/2$ and $S=1$, the sequence's spectral resolution is limited by the $T_1$ relaxation time of the electron spin.
On the other hand, the sequence's sensitivity is limited by $T_2$.
The main principle of 5-pulse ESEEM \cite{kasumaj20085} is to create entanglement (via, e.g., Hahn-echo or CPMG block) before the free evolution of the nuclear spins and then to apply a second correlating sequence afterwards.
In other words, the initial density matrix for the Hahn Echo sequence is modified so that entanglement already exists in the electron polarization terms of the density matrix, thus limited by $T_1$ time.
Fig \ref{fig:seqs} shows the sequence, which can be interpreted as two Hahn Echos separated by a long free evolution $T\gg T_2^\ast$ such that the coherence of the electron spin vanishes. The analytical formula for this sequence is provided in the appendix \ref{app:5pESEEM}.
During the free evolution time, each nuclear spin oscillates with one of two resonance frequencies obtained from Eq. \ref{eq:H}.

\label{sec:discussion}
\begin{figure*}
	\begin{center}
		\includegraphics[width=1.0\textwidth]{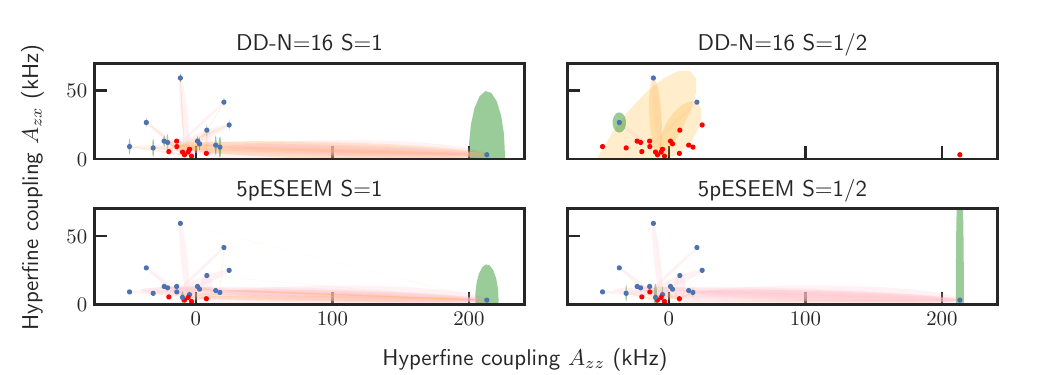}
		\caption{Visualisation of the CramerRao bound for various protocols on an existing nuclear spin dataset from \cite{abobeih2019atomic}}
		\label{fig:cov}
	\end{center}
\end{figure*}
Fig. \ref{fig:ddeseem} shows the Fourier transform of the signal.
Zooming into the area close to the Larmor frequency reveals the weakly coupled nuclei.
Each nuclear spin has two peaks in the spectrum referring to $\omega_a$ and $\omega_b$. Expanding the resonance frequencies up to first order with respect to hyperfine coupling shows that nuclear spins appear in order of their coupling strength, making their characterization rather straightforward.
To observe the weakly coupled nuclei more pronouncedly, one can use the DD-ESEEM sequence by adding more $\pi$ pulses in the entangling periods.
To distinguish other nuclei, extra measurements are required.
One method is to use two-dimensional hyperfine correlation spectroscopy (2D Hyscore) \cite{vorobyov2022addressing}.
However, this method is rather time-consuming as two time variables have to be swept.
Optimum time sampling techniques might be used to optimize the timing of 2D sampling.
Here we suggest an alternative way, which is conventionally used in ESR, i.e., sweeping the $\tau$ parameter and keeping track of the frequency amplitude as a function of $\tau$.
Two frequencies that belong to the same nucleus are correlated because the modulation depth oscillates with blind spot terms of both frequencies $\sin^2(\frac{\omega_0 \tau}{2}) \sin^2(\frac{\omega_1 \tau}{2})$.
Fig. \ref{fig:4} shows two frequencies going to bright and blind spots simultaneously.
Hence, by taking a two-dimensional correlation of the spectrum, one can deduce which two frequencies are correlated and belong to the same nucleus.

To simulate the realistic experimental situation, we model the electron spin state as projected to 0 and 1 with a binomial distribution where the probability is determined by the analytical expressions. Then, we assume the bright (dark) state emits 3 (0.1) photons on average with a Poissonian distribution. We repeat the measurement at each point for 10,000 times to reduce classical photon shot noise.

To estimate the number of nuclear spins that could be identified, we consider the duration of the free evolution time and the choice of inter-pulse timing $\tau$.
For a realistic scenario with $T_1 = 1$ s and $T_2 = 100 \mu s$, and sweeping the relevant parameter in 1000 steps, we can estimate the Fisher information matrix for various protocols.
We will approach the system with a moderate number of pulses, say $N=16$, for the dynamical decoupling sequence, and the simplest 5-pulse ESEEM for the correlation protocol for simplicity of calculation, as analytical expressions are available in that case.

Fig. \ref{fig:cov} illustrates the manifestation of the correlation protocol for a spin 1/2 system in terms of the number of nuclear spins one could estimate with each of the protocols.
The dots in the $A_{zz}, A_{zx}$ axes represent the points, with an area bounded by the Cramér-Rao bound around each point.
A nuclear spin is considered detectable (blue points in Fig. \ref{fig:cov}) if the uncertainty of one of the two hyperfine parameters is less than both the absolute value of that parameter and the distance to the closest hyperfine coupling.
Otherwise, it is considered non-identified (shown as red points). 
Additionally, covariance between various parameters is calculated from the off-diagonal elements in the inverse Fisher information matrix.
To illustrate them, an ellipse for each covariance parameter is plotted (colored orange and pink), with two nuclei as the vertices of the ellipse and the small axis size as the Cramér-Rao bound.
This means that in the case of zero covariance between two nuclei, this ellipse turns into a line.
To keep the plot readable, for each nuclear spin, only the ellipse with the largest covariance value is plotted.
These covariances mostly show the crosstalk between different parameters in the presented data and are more strongly present in the DD S=1/2 case. 
The idea behind the ellipse representation is that when the covariance is larger than the self-variance, the spins become indistinguishable.
While this crosstalk is not the limiting factor for informational approach, it might be important when considering a realistic estimator.
It is important to note that this is not strictly the same condition as in experimental identification and is idealized.
We only consider the available information in the plot, assuming that an optimum estimator for the extraction of that information already exists.
In reality, one has to additionally assume a non-ideal estimator procedure, which involves the extraction of the hyperfine parameter values from the raw data. But this is the subject of further work.
As a result, we see that while for the DD method, a spin 1/2 case for N=16 pulses allows for the identification of only 3 out of 23 nuclear spins, the 5-pulse ESEEM method works with similar success for both S=1 and S=1/2 systems. In total, it is capable of detecting 17-18 out of 23 spins within the measurement time constrain. This could be further boosted by using the DD-ESEEM method with multiple pulses in each sensing block analogous to Fig. \ref{fig:ddeseem}.

\section{Discussion}

In this work, we conducted a theoretical and numerical comparison of various ESEEM methods for characterizing the nuclear spin clusters around S=1 and S=1/2 type of defects.
We found that for S=1 systems, modified spin echo sequences such as CPMG and XY-N are most suited for single nuclear spin qubit spectroscopy, but for S=1/2 systems, their performance is limited. On the other hand, correlation-type sequences like 3 and 5 pulse ESEEM show more potential for characterizing a diluted nuclear spin bath and perform at least as well as in the S=1 case.
We believe that these methods hold significant potential for preliminary screening of nuclear spin clusters for S=1/2 potential qubit candidate systems, such as G-IV defects in diamond.
Additionally, our method does not require strong microwave pulse fields since the pulses do not need to cope with the nuclear spin Larmor frequency.
While we explore various sensing methods in this work, we did not delve into estimator performance.
We believe that this aspect should be considered in conjunction with adaptive and optimal strategies for controlling the experimental parameters, such as Bayesian Optimal Experimental Design or Machine Learning, to increase the efficiency of estimation.
This could further enhance the characterization capabilities of these methods in practical experimental scenarios.

\section*{Acknowledgments}
We acknowledge financial support by European Union's Horizon 2020 research and innovation program ASTERIQS under grant No. 820394 as well as Spinus, Federal Ministry of Education and Research (BMBF) project MiLiQuant and Quamapolis, Spinning and QR.X the DFG (FOR 2724, INST 41/1109-1 FUGG), the Max Planck Society, and the Volkswagentiftung. M.Z. thanks Max Planck School of Photonics for financial support.

\appendix
\section{Sequences}
\label{app:seqs}
The details of mentioned sequences in the main text can be found in fig \ref{fig:seqs}:
\begin{figure}
	\begin{center}
		\includegraphics[width=0.9\columnwidth]{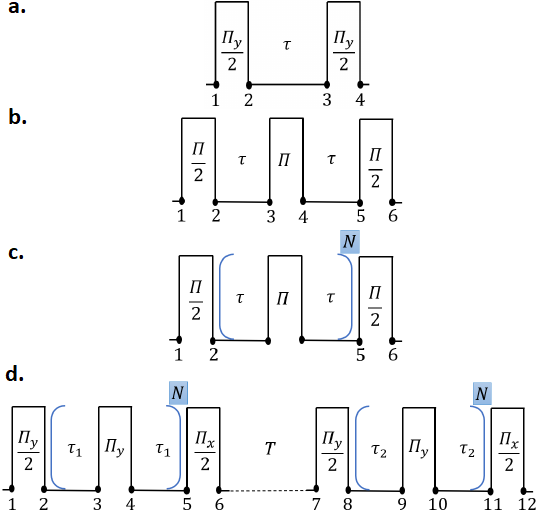}
		\caption{Pulse sequences a. Ramsey b. Hahn Echo c. Dynamical Decoupling d. DDESEEM}
		\label{fig:seqs}
	\end{center}
\end{figure}

\section{5pESEEM formula}
\label{app:5pESEEM} 

To define the unitary operators, let's introduce three sets of operators representing the evolution during the sequence:
1. The evolution during the first Hahn-echo is given by the 
V operators 2. The middle free evolution is represented by the 
F operators 3. The second Hahn-echo is governed by the W operators:
\begin{gather}
	V_0 = U_0 (\tau_1) U_1 (\tau_1),\,\,  V_1 = U_1 (\tau_1) U_0 (\tau_1) \\
	F_0 = U_0 (T), \,\, F_1 = U_1 (T) U_0 \\
	W_0 = U_0 (\tau_2) U_1 (\tau_2), \,\, W_1 = U_1 (\tau_2) U_0 (\tau_2)
\end{gather}
These operators define the trajectory of the electron spin during the 5-pulse ESEEM sequence.
The signal can be obtained from four different types of trajectories that the electron spin can take. We can express every unitary operator in terms of a rotation angle, axis, and Pauli matrices $\vec{\sigma}$. Denoting the unitary operator $U$ as $U = \exp(-i \theta_U \hat{n}_U\cdot\vec{\sigma}) = M\cos(\theta_U) \mathbf{I}- i \sin(\theta_U) \hat{n}_U\cdot\vec{\sigma}$. The signal for 5-pulse ESEEM can be obtained as follows:
\begin{gather}
		\expval{\sigma_z}_{5p} =\frac{1}{4}\Big( \prod_{j=1}^n [\cos(\theta_{W_1 F_1 V_0 V_1^\dagger F_1^\dagger W_0^\dagger})]^{(j)}  \notag \\
		-\prod_{j=1}^n [\cos(\theta_{W_1 F_1 V_1 V_0^\dagger F_1^\dagger W_0^\dagger})]^{(j)}  \notag \\
		+ \prod_{j=1}^n [\cos(\theta_{W_1 F_0 V_0 V_1^\dagger F_0^\dagger W_0^\dagger})]^{(j)}  \notag \\
		-\prod_{j=1}^n [\cos(\theta_{W_1 F_0 V_1 V_0^\dagger F_0^\dagger W_0^\dagger})]^{(j)} \Big) \textcolor{gray}{e^{-\frac{2\tau_1+2\tau_2}{T_2}- \frac{T}{T_1}}}
\end{gather}
Multiplying the matrices and finding the rotation angles gives the expectation value of $\sigma_z$ for 5pESEEM sequence as it can be found in this article \cite{kasumaj20085}:
\begin{gather}
	\expval{\sigma_z}_{5p} = \frac{1}{4}\Big( \prod_{j=1}^n E_{\alpha_+}^{(j)} - \prod_{j=1}^n E_{\alpha_-}^{(j)} + \prod_{j=1}^n E_{\beta_+}^{(j)}- \prod_{j=1}^n E_{\beta_-}^{(j)}\Big)
\end{gather}
Where each term can be calculated as follows:
\small   
\begin{gather} 
	E_{\alpha_\pm}^{(k)} = E_{2p}(\tau_1) E_{2p}(\tau_2) \mp B \Big(-4k^2C_\alpha \notag \\
	+4k\cos^4(\eta) \cos(\omega_\alpha T + \phi_{\alpha_+} +\phi_{\beta_+}) 
	+2 k^2 \cos(\phi_{\beta_-}) \cos(\omega_\alpha T+\phi_{\alpha_+}) \notag \\
	+ 4k \sin^4(\eta) \cos(\omega_\alpha T +\phi_{\alpha_+} -\phi_{\beta_+} )\Big)
	\label{eq:eseem}
\end{gather}
\normalsize
$ E_{2p} $ is 2-pulse ESEEM sequence signal:
\begin{gather}
	E_{2p}(t)  = (1 - \frac{k}{2}) + \frac{k}{2} \Big(\cos(\omega_\alpha t)+\cos(\omega_\beta t) \notag \\
	-\frac{1}{2} \cos(\omega_{-}t)  -\frac{1}{2} \cos(\omega_{+}t)   \Big)
\end{gather}
With $\omega_{\pm} = \omega_\alpha \pm \omega_\beta$,  B is the blind spot term and $C_\alpha$ is a constant term:
\begin{gather}
	B = \sin(\frac{\omega_\alpha \tau_1}{2}) \sin(\frac{\omega_\alpha \tau_2}{2}) \sin(\frac{\omega_\beta \tau_1}{2}) \sin(\frac{\omega_\beta \tau_2}{2})
	\label{eq:BS}\\
	C_\alpha =\cos(\frac{\omega_\alpha \tau_1}{2}) \cos(\frac{\omega_\alpha \tau_2}{2}) \sin(\frac{\omega_\beta \tau_1}{2}) \sin(\frac{\omega_\beta \tau_2}{2}) \label{eq:Cterms}
\end{gather}
The resonance frequencies are $\omega_{\alpha(\beta)} = \sqrt{(\omega_L +s_{0(1)} A_{zz} )^2 +(s_{0(1)} A_{zx})^2}$. The quantization axis of nuclear spins is tilted by $\eta_{\alpha(\beta)} = \arctan(\frac{s_{0(1)}A_{zx}}{\omega_L+ s_{0(1)}A_{zz}})$, which gives the parameter $\eta=\frac{\eta_\alpha - \eta_\beta}{2}$. The modulation depth of each nuclear spin is $k=\sin^2(2\eta)=(\frac{(s_1-s_0)\omega_L A_{zx}}{\omega_\alpha \omega_\beta})^2$, the phase shifts are $\phi_{\alpha_\pm} = \frac{\omega_\alpha (\tau_1 \pm \tau_2)}{2}$ and $\phi_{\beta_\pm} = \frac{\omega_\beta (\tau_1 \pm \tau_2)}{2}$. The two expression for the $\beta$ pathways can be obtained by exchanging $\alpha$ and $\beta$ in equation \ref{eq:eseem}, \ref{eq:BS}, and \ref{eq:Cterms}.\\

As demonstrated, the signal from the nuclear spin register arises from the multiplication of signals from individual nuclear spins.
Consequently, if the modulation depth is not low (indicative of low magnetic field conditions), higher-order frequencies will manifest in the spectrum.
Thus, the product rule gives rise to inter-nuclear peaks at multi-quantum frequencies, which can represent sums or subtractions of single quantum frequencies from various nuclei.
However, these multiple quantum resonance peaks are informative for electron spin one-half systems because it enables deduction that two peaks added or subtracted belong to the same electron spin manifold, allowing for determination of relative phase of nuclei.
Another consequence of the product rule is the cross-suppression effect, where the presence of strongly coupled nuclei suppresses the amplitude of weakly coupled nuclei, while weakly coupled ones do not suppress the amplitude of strongly coupled nuclei \cite{stoll2005peak}. However, assuming a relatively high Larmor frequency or low modulation depth, both of these effects will vanish as the product rule can be approximated by a summation rule:\\
\begin{gather}
	E_\alpha = \prod_{j=1}^n E_{\alpha_+}^{(j)} - \prod_{j=1}^n E_{\alpha_-}^{(j)}  \notag \\ 
	\approx \sum_{j=1}^n \Big[-8Bk\cos^4(\eta) \cos(\omega_\alpha T + \phi_{\alpha_+}+ \phi_{\beta_+}) \Big]^{(j)}
\end{gather}

The blind spot term shows that how this sequence can be engineered to increase or reduce the signal amplitude of one nuclear spin from the spectrum. The bright and blind spots of a frequency in the spectrum are as follows:
\begin{align}
	\text{Blind Spots:	} \tau = even \,\, \frac{\pi}{\omega}\\
	\text{Bright Spots:	} \tau = odd \,\, \frac{\pi}{\omega}
\end{align}
The blind spots does not depend only on one frequency but on the nuclear spin. It means if one resonant frequency is blinded, the other resonant frequency and all the multiple quantum resonances also vanishes. This can be used as a manifestation that which two peaks are from one nuclei. Also, This is very helpful especially in the presence of strongly coupled nuclear spin that suppress other nuclei. By sweeping $\tau_1$ and $\tau_2$, one can go through different bright and blind spots of each nuclear spin, observe which two peaks are correlated.

\section{DD analysis for bispecies systems}
\label{app:bispecies}
Consider a single color center that is surrounded by two nuclear spin species. Each species has a distinct bath, which is processing with its corresponding Larmor frequency. Assuming large Larmor frequency with respect to hyperfine couplings, one can write Eq. \ref{eq:approx_DD} for two bath ($\omega_0 \approx \omega_1 \approx \omega_L$) as follows:
\begin{gather}
	\expval{\sigma_z}_{bath} \approx 1 -2 k_1^2 \sin^4(\frac{\omega_L^{(1)} \tau}{2}) -2 k_2^2 \sin^4(\frac{\omega_L^{(2)} \tau}{2})
\end{gather}
This means that two baths interfere, and the periodicity of the total bath is not straightforward any more. Hence, if there is a narrow peak next in DD signal, it can be attributed to both of the baths. To clarify the type of nuclear spins, one has to perform further experiments, e.g., in various external magnetic fields.

\begin{thebibliography}{24}%
\makeatletter
\providecommand \@ifxundefined [1]{%
 \@ifx{#1\undefined}
}%
\providecommand \@ifnum [1]{%
 \ifnum #1\expandafter \@firstoftwo
 \else \expandafter \@secondoftwo
 \fi
}%
\providecommand \@ifx [1]{%
 \ifx #1\expandafter \@firstoftwo
 \else \expandafter \@secondoftwo
 \fi
}%
\providecommand \natexlab [1]{#1}%
\providecommand \enquote  [1]{``#1''}%
\providecommand \bibnamefont  [1]{#1}%
\providecommand \bibfnamefont [1]{#1}%
\providecommand \citenamefont [1]{#1}%
\providecommand \href@noop [0]{\@secondoftwo}%
\providecommand \href [0]{\begingroup \@sanitize@url \@href}%
\providecommand \@href[1]{\@@startlink{#1}\@@href}%
\providecommand \@@href[1]{\endgroup#1\@@endlink}%
\providecommand \@sanitize@url [0]{\catcode `\\12\catcode `\$12\catcode
  `\&12\catcode `\#12\catcode `\^12\catcode `\_12\catcode `\%12\relax}%
\providecommand \@@startlink[1]{}%
\providecommand \@@endlink[0]{}%
\providecommand \url  [0]{\begingroup\@sanitize@url \@url }%
\providecommand \@url [1]{\endgroup\@href {#1}{\urlprefix }}%
\providecommand \urlprefix  [0]{URL }%
\providecommand \Eprint [0]{\href }%
\providecommand \doibase [0]{https://doi.org/}%
\providecommand \selectlanguage [0]{\@gobble}%
\providecommand \bibinfo  [0]{\@secondoftwo}%
\providecommand \bibfield  [0]{\@secondoftwo}%
\providecommand \translation [1]{[#1]}%
\providecommand \BibitemOpen [0]{}%
\providecommand \bibitemStop [0]{}%
\providecommand \bibitemNoStop [0]{.\EOS\space}%
\providecommand \EOS [0]{\spacefactor3000\relax}%
\providecommand \BibitemShut  [1]{\csname bibitem#1\endcsname}%
\let\auto@bib@innerbib\@empty
\bibitem [{\citenamefont {Awschalom}\ \emph {et~al.}(2018)\citenamefont
  {Awschalom}, \citenamefont {Hanson}, \citenamefont {Wrachtrup},\ and\
  \citenamefont {Zhou}}]{awschalom2018quantum}%
  \BibitemOpen
  \bibfield  {author} {\bibinfo {author} {\bibfnamefont {D.~D.}\ \bibnamefont
  {Awschalom}}, \bibinfo {author} {\bibfnamefont {R.}~\bibnamefont {Hanson}},
  \bibinfo {author} {\bibfnamefont {J.}~\bibnamefont {Wrachtrup}},\ and\
  \bibinfo {author} {\bibfnamefont {B.~B.}\ \bibnamefont {Zhou}},\ }\bibfield
  {title} {\bibinfo {title} {Quantum technologies with optically interfaced
  solid-state spins},\ }\href@noop {} {\bibfield  {journal} {\bibinfo
  {journal} {Nature Photonics}\ }\textbf {\bibinfo {volume} {12}},\ \bibinfo
  {pages} {516} (\bibinfo {year} {2018})}\BibitemShut {NoStop}%
\bibitem [{\citenamefont {Pompili}\ \emph {et~al.}(2021)\citenamefont
  {Pompili}, \citenamefont {Hermans}, \citenamefont {Baier}, \citenamefont
  {Beukers}, \citenamefont {Humphreys}, \citenamefont {Schouten}, \citenamefont
  {Vermeulen}, \citenamefont {Tiggelman}, \citenamefont {dos Santos~Martins},
  \citenamefont {Dirkse} \emph {et~al.}}]{pompili2021realization}%
  \BibitemOpen
  \bibfield  {author} {\bibinfo {author} {\bibfnamefont {M.}~\bibnamefont
  {Pompili}}, \bibinfo {author} {\bibfnamefont {S.~L.}\ \bibnamefont
  {Hermans}}, \bibinfo {author} {\bibfnamefont {S.}~\bibnamefont {Baier}},
  \bibinfo {author} {\bibfnamefont {H.~K.}\ \bibnamefont {Beukers}}, \bibinfo
  {author} {\bibfnamefont {P.~C.}\ \bibnamefont {Humphreys}}, \bibinfo {author}
  {\bibfnamefont {R.~N.}\ \bibnamefont {Schouten}}, \bibinfo {author}
  {\bibfnamefont {R.~F.}\ \bibnamefont {Vermeulen}}, \bibinfo {author}
  {\bibfnamefont {M.~J.}\ \bibnamefont {Tiggelman}}, \bibinfo {author}
  {\bibfnamefont {L.}~\bibnamefont {dos Santos~Martins}}, \bibinfo {author}
  {\bibfnamefont {B.}~\bibnamefont {Dirkse}}, \emph {et~al.},\ }\bibfield
  {title} {\bibinfo {title} {Realization of a multinode quantum network of
  remote solid-state qubits},\ }\href@noop {} {\bibfield  {journal} {\bibinfo
  {journal} {Science}\ }\textbf {\bibinfo {volume} {372}},\ \bibinfo {pages}
  {259} (\bibinfo {year} {2021})}\BibitemShut {NoStop}%
\bibitem [{\citenamefont {Degen}\ \emph {et~al.}(2017)\citenamefont {Degen},
  \citenamefont {Reinhard},\ and\ \citenamefont
  {Cappellaro}}]{degen2017quantum}%
  \BibitemOpen
  \bibfield  {author} {\bibinfo {author} {\bibfnamefont {C.~L.}\ \bibnamefont
  {Degen}}, \bibinfo {author} {\bibfnamefont {F.}~\bibnamefont {Reinhard}},\
  and\ \bibinfo {author} {\bibfnamefont {P.}~\bibnamefont {Cappellaro}},\
  }\bibfield  {title} {\bibinfo {title} {Quantum sensing},\ }\href@noop {}
  {\bibfield  {journal} {\bibinfo  {journal} {Reviews of modern physics}\
  }\textbf {\bibinfo {volume} {89}},\ \bibinfo {pages} {035002} (\bibinfo
  {year} {2017})}\BibitemShut {NoStop}%
\bibitem [{\citenamefont {Meinel}\ \emph {et~al.}(2023)\citenamefont {Meinel},
  \citenamefont {Kwon}, \citenamefont {Maier}, \citenamefont {Dasari},
  \citenamefont {Sumiya}, \citenamefont {Onoda}, \citenamefont {Isoya},
  \citenamefont {Vorobyov},\ and\ \citenamefont {Wrachtrup}}]{meinel2023high}%
  \BibitemOpen
  \bibfield  {author} {\bibinfo {author} {\bibfnamefont {J.}~\bibnamefont
  {Meinel}}, \bibinfo {author} {\bibfnamefont {M.}~\bibnamefont {Kwon}},
  \bibinfo {author} {\bibfnamefont {R.}~\bibnamefont {Maier}}, \bibinfo
  {author} {\bibfnamefont {D.}~\bibnamefont {Dasari}}, \bibinfo {author}
  {\bibfnamefont {H.}~\bibnamefont {Sumiya}}, \bibinfo {author} {\bibfnamefont
  {S.}~\bibnamefont {Onoda}}, \bibinfo {author} {\bibfnamefont
  {J.}~\bibnamefont {Isoya}}, \bibinfo {author} {\bibfnamefont
  {V.}~\bibnamefont {Vorobyov}},\ and\ \bibinfo {author} {\bibfnamefont
  {J.}~\bibnamefont {Wrachtrup}},\ }\bibfield  {title} {\bibinfo {title}
  {High-resolution nanoscale nmr for arbitrary magnetic fields},\ }\href@noop
  {} {\bibfield  {journal} {\bibinfo  {journal} {Communications Physics}\
  }\textbf {\bibinfo {volume} {6}},\ \bibinfo {pages} {302} (\bibinfo {year}
  {2023})}\BibitemShut {NoStop}%
\bibitem [{\citenamefont {Bradley}\ \emph {et~al.}(2019)\citenamefont
  {Bradley}, \citenamefont {Randall}, \citenamefont {Abobeih}, \citenamefont
  {Berrevoets}, \citenamefont {Degen}, \citenamefont {Bakker}, \citenamefont
  {Markham}, \citenamefont {Twitchen},\ and\ \citenamefont
  {Taminiau}}]{bradley2019ten}%
  \BibitemOpen
  \bibfield  {author} {\bibinfo {author} {\bibfnamefont {C.~E.}\ \bibnamefont
  {Bradley}}, \bibinfo {author} {\bibfnamefont {J.}~\bibnamefont {Randall}},
  \bibinfo {author} {\bibfnamefont {M.~H.}\ \bibnamefont {Abobeih}}, \bibinfo
  {author} {\bibfnamefont {R.}~\bibnamefont {Berrevoets}}, \bibinfo {author}
  {\bibfnamefont {M.}~\bibnamefont {Degen}}, \bibinfo {author} {\bibfnamefont
  {M.~A.}\ \bibnamefont {Bakker}}, \bibinfo {author} {\bibfnamefont
  {M.}~\bibnamefont {Markham}}, \bibinfo {author} {\bibfnamefont
  {D.}~\bibnamefont {Twitchen}},\ and\ \bibinfo {author} {\bibfnamefont
  {T.~H.}\ \bibnamefont {Taminiau}},\ }\bibfield  {title} {\bibinfo {title} {A
  ten-qubit solid-state spin register with quantum memory up to one minute},\
  }\href@noop {} {\bibfield  {journal} {\bibinfo  {journal} {Physical Review
  X}\ }\textbf {\bibinfo {volume} {9}},\ \bibinfo {pages} {031045} (\bibinfo
  {year} {2019})}\BibitemShut {NoStop}%
\bibitem [{\citenamefont {Vorobyov}\ \emph {et~al.}(2021)\citenamefont
  {Vorobyov}, \citenamefont {Zaiser}, \citenamefont {Abt}, \citenamefont
  {Meinel}, \citenamefont {Dasari}, \citenamefont {Neumann},\ and\
  \citenamefont {Wrachtrup}}]{vorobyov2021quantum}%
  \BibitemOpen
  \bibfield  {author} {\bibinfo {author} {\bibfnamefont {V.}~\bibnamefont
  {Vorobyov}}, \bibinfo {author} {\bibfnamefont {S.}~\bibnamefont {Zaiser}},
  \bibinfo {author} {\bibfnamefont {N.}~\bibnamefont {Abt}}, \bibinfo {author}
  {\bibfnamefont {J.}~\bibnamefont {Meinel}}, \bibinfo {author} {\bibfnamefont
  {D.}~\bibnamefont {Dasari}}, \bibinfo {author} {\bibfnamefont
  {P.}~\bibnamefont {Neumann}},\ and\ \bibinfo {author} {\bibfnamefont
  {J.}~\bibnamefont {Wrachtrup}},\ }\bibfield  {title} {\bibinfo {title}
  {Quantum fourier transform for nanoscale quantum sensing},\ }\href@noop {}
  {\bibfield  {journal} {\bibinfo  {journal} {npj Quantum Information}\
  }\textbf {\bibinfo {volume} {7}},\ \bibinfo {pages} {124} (\bibinfo {year}
  {2021})}\BibitemShut {NoStop}%
\bibitem [{\citenamefont {Hesselmeier}\ \emph {et~al.}(2023)\citenamefont
  {Hesselmeier}, \citenamefont {Kuna}, \citenamefont {Tak{\'a}cs},
  \citenamefont {Iv{\'a}dy}, \citenamefont {Knolle}, \citenamefont {Ghezellou},
  \citenamefont {Ul-Hassan}, \citenamefont {Dasari}, \citenamefont {Kaiser},
  \citenamefont {Vorobyov} \emph {et~al.}}]{hesselmeier2023measuring}%
  \BibitemOpen
  \bibfield  {author} {\bibinfo {author} {\bibfnamefont {E.}~\bibnamefont
  {Hesselmeier}}, \bibinfo {author} {\bibfnamefont {P.}~\bibnamefont {Kuna}},
  \bibinfo {author} {\bibfnamefont {I.}~\bibnamefont {Tak{\'a}cs}}, \bibinfo
  {author} {\bibfnamefont {V.}~\bibnamefont {Iv{\'a}dy}}, \bibinfo {author}
  {\bibfnamefont {W.}~\bibnamefont {Knolle}}, \bibinfo {author} {\bibfnamefont
  {M.}~\bibnamefont {Ghezellou}}, \bibinfo {author} {\bibfnamefont
  {J.}~\bibnamefont {Ul-Hassan}}, \bibinfo {author} {\bibfnamefont
  {D.}~\bibnamefont {Dasari}}, \bibinfo {author} {\bibfnamefont
  {F.}~\bibnamefont {Kaiser}}, \bibinfo {author} {\bibfnamefont
  {V.}~\bibnamefont {Vorobyov}}, \emph {et~al.},\ }\bibfield  {title} {\bibinfo
  {title} {Measuring nuclear spin qubits by qudit-enhanced spectroscopy in
  silicon carbide},\ }\href@noop {} {\bibfield  {journal} {\bibinfo  {journal}
  {arXiv preprint arXiv:2310.15557}\ } (\bibinfo {year} {2023})}\BibitemShut
  {NoStop}%
\bibitem [{\citenamefont {Hesselmeier}\ \emph {et~al.}(2024)\citenamefont
  {Hesselmeier}, \citenamefont {Kuna}, \citenamefont {Knolle}, \citenamefont
  {Kaiser}, \citenamefont {Son}, \citenamefont {Ghezellou}, \citenamefont
  {Ul-Hassan}, \citenamefont {Vorobyov},\ and\ \citenamefont
  {Wrachtrup}}]{hesselmeier2024high}%
  \BibitemOpen
  \bibfield  {author} {\bibinfo {author} {\bibfnamefont {E.}~\bibnamefont
  {Hesselmeier}}, \bibinfo {author} {\bibfnamefont {P.}~\bibnamefont {Kuna}},
  \bibinfo {author} {\bibfnamefont {W.}~\bibnamefont {Knolle}}, \bibinfo
  {author} {\bibfnamefont {F.}~\bibnamefont {Kaiser}}, \bibinfo {author}
  {\bibfnamefont {N.~T.}\ \bibnamefont {Son}}, \bibinfo {author} {\bibfnamefont
  {M.}~\bibnamefont {Ghezellou}}, \bibinfo {author} {\bibfnamefont
  {J.}~\bibnamefont {Ul-Hassan}}, \bibinfo {author} {\bibfnamefont
  {V.}~\bibnamefont {Vorobyov}},\ and\ \bibinfo {author} {\bibfnamefont
  {J.}~\bibnamefont {Wrachtrup}},\ }\bibfield  {title} {\bibinfo {title} {High
  fidelity optical readout of a nuclear spin qubit in silicon carbide},\
  }\href@noop {} {\bibfield  {journal} {\bibinfo  {journal} {arXiv preprint
  arXiv:2401.04465}\ } (\bibinfo {year} {2024})}\BibitemShut {NoStop}%
\bibitem [{\citenamefont {Reiner}\ \emph {et~al.}(2024)\citenamefont {Reiner},
  \citenamefont {Chung}, \citenamefont {Misha}, \citenamefont {Lehner},
  \citenamefont {Moehle}, \citenamefont {Poulos}, \citenamefont {Monir},
  \citenamefont {Charde}, \citenamefont {Macha}, \citenamefont {Kranz} \emph
  {et~al.}}]{reiner2024high}%
  \BibitemOpen
  \bibfield  {author} {\bibinfo {author} {\bibfnamefont {J.}~\bibnamefont
  {Reiner}}, \bibinfo {author} {\bibfnamefont {Y.}~\bibnamefont {Chung}},
  \bibinfo {author} {\bibfnamefont {S.}~\bibnamefont {Misha}}, \bibinfo
  {author} {\bibfnamefont {C.}~\bibnamefont {Lehner}}, \bibinfo {author}
  {\bibfnamefont {C.}~\bibnamefont {Moehle}}, \bibinfo {author} {\bibfnamefont
  {D.}~\bibnamefont {Poulos}}, \bibinfo {author} {\bibfnamefont
  {S.}~\bibnamefont {Monir}}, \bibinfo {author} {\bibfnamefont
  {K.}~\bibnamefont {Charde}}, \bibinfo {author} {\bibfnamefont
  {P.}~\bibnamefont {Macha}}, \bibinfo {author} {\bibfnamefont
  {L.}~\bibnamefont {Kranz}}, \emph {et~al.},\ }\bibfield  {title} {\bibinfo
  {title} {High-fidelity initialization and control of electron and nuclear
  spins in a four-qubit register},\ }\href@noop {} {\bibfield  {journal}
  {\bibinfo  {journal} {Nature Nanotechnology}\ ,\ \bibinfo {pages} {1}}
  (\bibinfo {year} {2024})}\BibitemShut {NoStop}%
\bibitem [{\citenamefont {Vorobyov}\ \emph {et~al.}(2023)\citenamefont
  {Vorobyov}, \citenamefont {Meinel}, \citenamefont {Sumiya}, \citenamefont
  {Onoda}, \citenamefont {Isoya}, \citenamefont {Gulinsky},\ and\ \citenamefont
  {Wrachtrup}}]{vorobyov2023transition}%
  \BibitemOpen
  \bibfield  {author} {\bibinfo {author} {\bibfnamefont {V.~V.}\ \bibnamefont
  {Vorobyov}}, \bibinfo {author} {\bibfnamefont {J.}~\bibnamefont {Meinel}},
  \bibinfo {author} {\bibfnamefont {H.}~\bibnamefont {Sumiya}}, \bibinfo
  {author} {\bibfnamefont {S.}~\bibnamefont {Onoda}}, \bibinfo {author}
  {\bibfnamefont {J.}~\bibnamefont {Isoya}}, \bibinfo {author} {\bibfnamefont
  {O.}~\bibnamefont {Gulinsky}},\ and\ \bibinfo {author} {\bibfnamefont
  {J.}~\bibnamefont {Wrachtrup}},\ }\bibfield  {title} {\bibinfo {title}
  {Transition from quantum to classical dynamics in weak measurements and
  reconstruction of quantum correlation},\ }\href@noop {} {\bibfield  {journal}
  {\bibinfo  {journal} {Physical Review A}\ }\textbf {\bibinfo {volume}
  {107}},\ \bibinfo {pages} {042212} (\bibinfo {year} {2023})}\BibitemShut
  {NoStop}%
\bibitem [{\citenamefont {Bradley}\ \emph {et~al.}(2022)\citenamefont
  {Bradley}, \citenamefont {de~Bone}, \citenamefont {M{\"o}ller}, \citenamefont
  {Baier}, \citenamefont {Degen}, \citenamefont {Loenen}, \citenamefont
  {Bartling}, \citenamefont {Markham}, \citenamefont {Twitchen}, \citenamefont
  {Hanson} \emph {et~al.}}]{bradley2022robust}%
  \BibitemOpen
  \bibfield  {author} {\bibinfo {author} {\bibfnamefont {C.}~\bibnamefont
  {Bradley}}, \bibinfo {author} {\bibfnamefont {S.}~\bibnamefont {de~Bone}},
  \bibinfo {author} {\bibfnamefont {P.}~\bibnamefont {M{\"o}ller}}, \bibinfo
  {author} {\bibfnamefont {S.}~\bibnamefont {Baier}}, \bibinfo {author}
  {\bibfnamefont {M.}~\bibnamefont {Degen}}, \bibinfo {author} {\bibfnamefont
  {S.}~\bibnamefont {Loenen}}, \bibinfo {author} {\bibfnamefont
  {H.}~\bibnamefont {Bartling}}, \bibinfo {author} {\bibfnamefont
  {M.}~\bibnamefont {Markham}}, \bibinfo {author} {\bibfnamefont
  {D.}~\bibnamefont {Twitchen}}, \bibinfo {author} {\bibfnamefont
  {R.}~\bibnamefont {Hanson}}, \emph {et~al.},\ }\bibfield  {title} {\bibinfo
  {title} {Robust quantum-network memory based on spin qubits in isotopically
  engineered diamond},\ }\href@noop {} {\bibfield  {journal} {\bibinfo
  {journal} {npj Quantum Information}\ }\textbf {\bibinfo {volume} {8}},\
  \bibinfo {pages} {122} (\bibinfo {year} {2022})}\BibitemShut {NoStop}%
\bibitem [{\citenamefont {Abobeih}\ \emph {et~al.}(2019)\citenamefont
  {Abobeih}, \citenamefont {Randall}, \citenamefont {Bradley}, \citenamefont
  {Bartling}, \citenamefont {Bakker}, \citenamefont {Degen}, \citenamefont
  {Markham}, \citenamefont {Twitchen},\ and\ \citenamefont
  {Taminiau}}]{abobeih2019atomic}%
  \BibitemOpen
  \bibfield  {author} {\bibinfo {author} {\bibfnamefont {M.}~\bibnamefont
  {Abobeih}}, \bibinfo {author} {\bibfnamefont {J.}~\bibnamefont {Randall}},
  \bibinfo {author} {\bibfnamefont {C.}~\bibnamefont {Bradley}}, \bibinfo
  {author} {\bibfnamefont {H.}~\bibnamefont {Bartling}}, \bibinfo {author}
  {\bibfnamefont {M.}~\bibnamefont {Bakker}}, \bibinfo {author} {\bibfnamefont
  {M.}~\bibnamefont {Degen}}, \bibinfo {author} {\bibfnamefont
  {M.}~\bibnamefont {Markham}}, \bibinfo {author} {\bibfnamefont
  {D.}~\bibnamefont {Twitchen}},\ and\ \bibinfo {author} {\bibfnamefont
  {T.}~\bibnamefont {Taminiau}},\ }\bibfield  {title} {\bibinfo {title}
  {Atomic-scale imaging of a 27-nuclear-spin cluster using a quantum sensor},\
  }\href@noop {} {\bibfield  {journal} {\bibinfo  {journal} {Nature}\ }\textbf
  {\bibinfo {volume} {576}},\ \bibinfo {pages} {411} (\bibinfo {year}
  {2019})}\BibitemShut {NoStop}%
\bibitem [{\citenamefont {van~de Stolpe}\ \emph {et~al.}(2023)\citenamefont
  {van~de Stolpe}, \citenamefont {Kwiatkowski}, \citenamefont {Bradley},
  \citenamefont {Randall}, \citenamefont {Breitweiser}, \citenamefont
  {Bassett}, \citenamefont {Markham}, \citenamefont {Twitchen},\ and\
  \citenamefont {Taminiau}}]{van2023mapping}%
  \BibitemOpen
  \bibfield  {author} {\bibinfo {author} {\bibfnamefont {G.}~\bibnamefont
  {van~de Stolpe}}, \bibinfo {author} {\bibfnamefont {D.}~\bibnamefont
  {Kwiatkowski}}, \bibinfo {author} {\bibfnamefont {C.}~\bibnamefont
  {Bradley}}, \bibinfo {author} {\bibfnamefont {J.}~\bibnamefont {Randall}},
  \bibinfo {author} {\bibfnamefont {S.}~\bibnamefont {Breitweiser}}, \bibinfo
  {author} {\bibfnamefont {L.}~\bibnamefont {Bassett}}, \bibinfo {author}
  {\bibfnamefont {M.}~\bibnamefont {Markham}}, \bibinfo {author} {\bibfnamefont
  {D.}~\bibnamefont {Twitchen}},\ and\ \bibinfo {author} {\bibfnamefont
  {T.}~\bibnamefont {Taminiau}},\ }\bibfield  {title} {\bibinfo {title}
  {Mapping a 50-spin-qubit network through correlated sensing},\ }\href@noop {}
  {\bibfield  {journal} {\bibinfo  {journal} {arXiv preprint arXiv:2307.06939}\
  } (\bibinfo {year} {2023})}\BibitemShut {NoStop}%
\bibitem [{\citenamefont {Dr{\'e}au}(2012)}]{Dreau:2012aa}%
  \BibitemOpen
  \bibfield  {author} {\bibinfo {author} {\bibfnamefont {A.}~\bibnamefont
  {Dr{\'e}au}},\ }\bibfield  {title} {\bibinfo {title} {High-resolution
  spectroscopy of single nv defects coupled with nearby $^{13}$c nuclear spins
  in diamond},\ }\bibfield  {journal} {\bibinfo  {journal} {Physical Review B}\
  }\textbf {\bibinfo {volume} {85}},\ \href
  {https://doi.org/10.1103/PhysRevB.85.134107} {10.1103/PhysRevB.85.134107}
  (\bibinfo {year} {2012})\BibitemShut {NoStop}%
\bibitem [{\citenamefont {Childress}\ \emph {et~al.}(2006)\citenamefont
  {Childress}, \citenamefont {Gurudev~Dutt}, \citenamefont {Taylor},
  \citenamefont {Zibrov}, \citenamefont {Jelezko}, \citenamefont {Wrachtrup},
  \citenamefont {Hemmer},\ and\ \citenamefont {Lukin}}]{childress2006coherent}%
  \BibitemOpen
  \bibfield  {author} {\bibinfo {author} {\bibfnamefont {L.}~\bibnamefont
  {Childress}}, \bibinfo {author} {\bibfnamefont {M.}~\bibnamefont
  {Gurudev~Dutt}}, \bibinfo {author} {\bibfnamefont {J.}~\bibnamefont
  {Taylor}}, \bibinfo {author} {\bibfnamefont {A.}~\bibnamefont {Zibrov}},
  \bibinfo {author} {\bibfnamefont {F.}~\bibnamefont {Jelezko}}, \bibinfo
  {author} {\bibfnamefont {J.}~\bibnamefont {Wrachtrup}}, \bibinfo {author}
  {\bibfnamefont {P.}~\bibnamefont {Hemmer}},\ and\ \bibinfo {author}
  {\bibfnamefont {M.}~\bibnamefont {Lukin}},\ }\bibfield  {title} {\bibinfo
  {title} {Coherent dynamics of coupled electron and nuclear spin qubits in
  diamond},\ }\href@noop {} {\bibfield  {journal} {\bibinfo  {journal}
  {Science}\ }\textbf {\bibinfo {volume} {314}},\ \bibinfo {pages} {281}
  (\bibinfo {year} {2006})}\BibitemShut {NoStop}%
\bibitem [{\citenamefont {Laraoui}\ \emph {et~al.}(2013)\citenamefont
  {Laraoui}, \citenamefont {Dolde}, \citenamefont {Burk}, \citenamefont
  {Reinhard}, \citenamefont {Wrachtrup},\ and\ \citenamefont
  {Meriles}}]{laraoui2013high}%
  \BibitemOpen
  \bibfield  {author} {\bibinfo {author} {\bibfnamefont {A.}~\bibnamefont
  {Laraoui}}, \bibinfo {author} {\bibfnamefont {F.}~\bibnamefont {Dolde}},
  \bibinfo {author} {\bibfnamefont {C.}~\bibnamefont {Burk}}, \bibinfo {author}
  {\bibfnamefont {F.}~\bibnamefont {Reinhard}}, \bibinfo {author}
  {\bibfnamefont {J.}~\bibnamefont {Wrachtrup}},\ and\ \bibinfo {author}
  {\bibfnamefont {C.~A.}\ \bibnamefont {Meriles}},\ }\bibfield  {title}
  {\bibinfo {title} {High-resolution correlation spectroscopy of 13c spins near
  a nitrogen-vacancy centre in diamond},\ }\href@noop {} {\bibfield  {journal}
  {\bibinfo  {journal} {Nature communications}\ }\textbf {\bibinfo {volume}
  {4}},\ \bibinfo {pages} {1651} (\bibinfo {year} {2013})}\BibitemShut
  {NoStop}%
\bibitem [{\citenamefont {Vorobyov}\ \emph {et~al.}(2022)\citenamefont
  {Vorobyov}, \citenamefont {Javadzade}, \citenamefont {Joliffe}, \citenamefont
  {Kaiser},\ and\ \citenamefont {Wrachtrup}}]{vorobyov2022addressing}%
  \BibitemOpen
  \bibfield  {author} {\bibinfo {author} {\bibfnamefont {V.}~\bibnamefont
  {Vorobyov}}, \bibinfo {author} {\bibfnamefont {J.}~\bibnamefont {Javadzade}},
  \bibinfo {author} {\bibfnamefont {M.}~\bibnamefont {Joliffe}}, \bibinfo
  {author} {\bibfnamefont {F.}~\bibnamefont {Kaiser}},\ and\ \bibinfo {author}
  {\bibfnamefont {J.}~\bibnamefont {Wrachtrup}},\ }\bibfield  {title} {\bibinfo
  {title} {Addressing single nuclear spins quantum memories by a central
  electron spin},\ }\href@noop {} {\bibfield  {journal} {\bibinfo  {journal}
  {Applied Magnetic Resonance}\ }\textbf {\bibinfo {volume} {53}},\ \bibinfo
  {pages} {1317} (\bibinfo {year} {2022})}\BibitemShut {NoStop}%
\bibitem [{\citenamefont {Taminiau}\ \emph {et~al.}(2012)\citenamefont
  {Taminiau}, \citenamefont {Wagenaar}, \citenamefont {Van~der Sar},
  \citenamefont {Jelezko}, \citenamefont {Dobrovitski},\ and\ \citenamefont
  {Hanson}}]{taminiau2012detection}%
  \BibitemOpen
  \bibfield  {author} {\bibinfo {author} {\bibfnamefont {T.}~\bibnamefont
  {Taminiau}}, \bibinfo {author} {\bibfnamefont {J.}~\bibnamefont {Wagenaar}},
  \bibinfo {author} {\bibfnamefont {T.}~\bibnamefont {Van~der Sar}}, \bibinfo
  {author} {\bibfnamefont {F.}~\bibnamefont {Jelezko}}, \bibinfo {author}
  {\bibfnamefont {V.~V.}\ \bibnamefont {Dobrovitski}},\ and\ \bibinfo {author}
  {\bibfnamefont {R.}~\bibnamefont {Hanson}},\ }\bibfield  {title} {\bibinfo
  {title} {Detection and control of individual nuclear spins using a weakly
  coupled electron spin},\ }\href@noop {} {\bibfield  {journal} {\bibinfo
  {journal} {Physical review letters}\ }\textbf {\bibinfo {volume} {109}},\
  \bibinfo {pages} {137602} (\bibinfo {year} {2012})}\BibitemShut {NoStop}%
\bibitem [{\citenamefont {Zhao}(2014)}]{Zhao:2014aa}%
  \BibitemOpen
  \bibfield  {author} {\bibinfo {author} {\bibfnamefont {N.}~\bibnamefont
  {Zhao}},\ }\bibfield  {title} {\bibinfo {title} {Dynamical decoupling design
  for identifying weakly coupled nuclear spins in a bath},\ }\bibfield
  {journal} {\bibinfo  {journal} {Physical Review A}\ }\textbf {\bibinfo
  {volume} {90}},\ \href {https://doi.org/10.1103/PhysRevA.90.032319}
  {10.1103/PhysRevA.90.032319} (\bibinfo {year} {2014})\BibitemShut {NoStop}%
\bibitem [{\citenamefont {Neumann}\ \emph {et~al.}(2010)\citenamefont
  {Neumann}, \citenamefont {Beck}, \citenamefont {Steiner}, \citenamefont
  {Rempp}, \citenamefont {Fedder}, \citenamefont {Hemmer}, \citenamefont
  {Wrachtrup},\ and\ \citenamefont {Jelezko}}]{neumann2010single}%
  \BibitemOpen
  \bibfield  {author} {\bibinfo {author} {\bibfnamefont {P.}~\bibnamefont
  {Neumann}}, \bibinfo {author} {\bibfnamefont {J.}~\bibnamefont {Beck}},
  \bibinfo {author} {\bibfnamefont {M.}~\bibnamefont {Steiner}}, \bibinfo
  {author} {\bibfnamefont {F.}~\bibnamefont {Rempp}}, \bibinfo {author}
  {\bibfnamefont {H.}~\bibnamefont {Fedder}}, \bibinfo {author} {\bibfnamefont
  {P.~R.}\ \bibnamefont {Hemmer}}, \bibinfo {author} {\bibfnamefont
  {J.}~\bibnamefont {Wrachtrup}},\ and\ \bibinfo {author} {\bibfnamefont
  {F.}~\bibnamefont {Jelezko}},\ }\bibfield  {title} {\bibinfo {title}
  {Single-shot readout of a single nuclear spin},\ }\href@noop {} {\bibfield
  {journal} {\bibinfo  {journal} {science}\ }\textbf {\bibinfo {volume}
  {329}},\ \bibinfo {pages} {542} (\bibinfo {year} {2010})}\BibitemShut
  {NoStop}%
\bibitem [{\citenamefont {Bradley}(2019)}]{Bradley:2019aa}%
  \BibitemOpen
  \bibfield  {author} {\bibinfo {author} {\bibfnamefont {C.}~\bibnamefont
  {Bradley}},\ }\bibfield  {title} {\bibinfo {title} {A ten-qubit solid-state
  spin register with quantum memory up to one minute},\ }\bibfield  {journal}
  {\bibinfo  {journal} {Physical Review X}\ }\textbf {\bibinfo {volume} {9}},\
  \href {https://doi.org/10.1103/PhysRevX.9.031045} {10.1103/PhysRevX.9.031045}
  (\bibinfo {year} {2019})\BibitemShut {NoStop}%
\bibitem [{\citenamefont {Zaiser}\ \emph {et~al.}(2016)\citenamefont {Zaiser},
  \citenamefont {Rendler}, \citenamefont {Jakobi}, \citenamefont {Wolf},
  \citenamefont {Lee}, \citenamefont {Wagner}, \citenamefont {Bergholm},
  \citenamefont {Schulte-Herbr{\"u}ggen}, \citenamefont {Neumann},\ and\
  \citenamefont {Wrachtrup}}]{zaiser2016enhancing}%
  \BibitemOpen
  \bibfield  {author} {\bibinfo {author} {\bibfnamefont {S.}~\bibnamefont
  {Zaiser}}, \bibinfo {author} {\bibfnamefont {T.}~\bibnamefont {Rendler}},
  \bibinfo {author} {\bibfnamefont {I.}~\bibnamefont {Jakobi}}, \bibinfo
  {author} {\bibfnamefont {T.}~\bibnamefont {Wolf}}, \bibinfo {author}
  {\bibfnamefont {S.-Y.}\ \bibnamefont {Lee}}, \bibinfo {author} {\bibfnamefont
  {S.}~\bibnamefont {Wagner}}, \bibinfo {author} {\bibfnamefont
  {V.}~\bibnamefont {Bergholm}}, \bibinfo {author} {\bibfnamefont
  {T.}~\bibnamefont {Schulte-Herbr{\"u}ggen}}, \bibinfo {author} {\bibfnamefont
  {P.}~\bibnamefont {Neumann}},\ and\ \bibinfo {author} {\bibfnamefont
  {J.}~\bibnamefont {Wrachtrup}},\ }\bibfield  {title} {\bibinfo {title}
  {Enhancing quantum sensing sensitivity by a quantum memory},\ }\href@noop {}
  {\bibfield  {journal} {\bibinfo  {journal} {Nature communications}\ }\textbf
  {\bibinfo {volume} {7}},\ \bibinfo {pages} {12279} (\bibinfo {year}
  {2016})}\BibitemShut {NoStop}%
\bibitem [{\citenamefont {Kasumaj}\ and\ \citenamefont
  {Stoll}(2008)}]{kasumaj20085}%
  \BibitemOpen
  \bibfield  {author} {\bibinfo {author} {\bibfnamefont {B.}~\bibnamefont
  {Kasumaj}}\ and\ \bibinfo {author} {\bibfnamefont {S.}~\bibnamefont
  {Stoll}},\ }\bibfield  {title} {\bibinfo {title} {5-and 6-pulse electron spin
  echo envelope modulation (eseem) of multi-nuclear spin systems},\ }\href@noop
  {} {\bibfield  {journal} {\bibinfo  {journal} {Journal of Magnetic
  Resonance}\ }\textbf {\bibinfo {volume} {190}},\ \bibinfo {pages} {233}
  (\bibinfo {year} {2008})}\BibitemShut {NoStop}%
\bibitem [{\citenamefont {Stoll}\ \emph {et~al.}(2005)\citenamefont {Stoll},
  \citenamefont {Calle}, \citenamefont {Mitrikas},\ and\ \citenamefont
  {Schweiger}}]{stoll2005peak}%
  \BibitemOpen
  \bibfield  {author} {\bibinfo {author} {\bibfnamefont {S.}~\bibnamefont
  {Stoll}}, \bibinfo {author} {\bibfnamefont {C.}~\bibnamefont {Calle}},
  \bibinfo {author} {\bibfnamefont {G.}~\bibnamefont {Mitrikas}},\ and\
  \bibinfo {author} {\bibfnamefont {A.}~\bibnamefont {Schweiger}},\ }\bibfield
  {title} {\bibinfo {title} {Peak suppression in eseem spectra of multinuclear
  spin systems},\ }\href@noop {} {\bibfield  {journal} {\bibinfo  {journal}
  {Journal of Magnetic Resonance}\ }\textbf {\bibinfo {volume} {177}},\
  \bibinfo {pages} {93} (\bibinfo {year} {2005})}\BibitemShut {NoStop}%
\end{thebibliography}
%

\end{document}